# Quantitative phase imaging with molecular vibrational sensitivity

Miu Tamamitsu,[1] Keiichiro Toda,[1] Ryoichi Horisaki,[2,3] and Takuro Ideguchi[1,3,*]

[1]Department of Physics, The University of Tokyo, Tokyo 113-0033, Japan
[2]Graduate School of Information Science and Technology, Osaka University, Osaka 565-0871, Japan
[3]PRESTO, Japan Science and Technology Agency, Saitama 332-0012, Japan
*Corresponding author: ideguchi@phys.s.u-tokyo.ac.jp

**Abstract**
Quantitative phase imaging (QPI) quantifies the sample-specific optical-phase-delay enabling objective studies of optically-transparent specimens such as biological samples, but lacks chemical sensitivity limiting its application to morphology-based diagnosis. We present wide-field molecular-vibrational microscopy realized in the framework of QPI utilizing mid-infrared photothermal effect. Our technique provides mid-infrared spectroscopic performance comparable to that of a conventional infrared spectrometer in the molecular fingerprint region of 1,450 – 1,600 cm$^{-1}$ and realizes wide-field molecular imaging of silica-polystyrene beads mixture over 100 μm × 100 μm area at 1 frame per second with the spatial resolution of 430 nm and 2 - 3 orders of magnitude lower fluence of ~10 pJ/μm$^2$ compared to other high-speed label-free molecular imaging methods, reducing photodamages to the sample. With a high-energy mid-infrared pulse source, our technique could enable high-speed, label-free, simultaneous and in-situ acquisition of quantitative morphology and molecular-vibrational contrast, providing new insights for studies of optically-transparent complex dynamics.

**Introduction**
Quantitative phase imaging (QPI) has become a valuable tool for studying transparent specimens such as biological cells and tissues. Unlike phase-contrast or differential-interference-contrast microscopy, QPI provides the quantitative measure of the optical-path-length delay introduced by the specimen in terms of the optical-phase delay at each spatial point of the field of view by means of interferometry [1] or phase retrieval of a diffraction pattern of the electromagnetic field [2] with a nanometer resolution, which allows for high-contrast and objective studies of microscopic specimens. A significant advantage of QPI is its wide-field label-free measurement capability of optically-transparent morphology, enabling high-speed imaging limited by the image sensor's frame rate while reducing optical and/or chemical damages to the sample which are troublesome in other imaging techniques such as staining, fluorescence [3] and Raman [4] imaging. However, QPI lacks chemical sensitivity which makes it difficult to determine the molecular compositions of the sample and limits its application to morphology-based image diagnosis [5].

Here, we propose and demonstrate wide-field, label-free molecular-vibrational (MV) microscopy realized in the framework of QPI utilizing mid-infrared (MIR) photothermal effect. The fundamental vibrational frequencies of various molecular species are found in the MIR electromagnetic region, hence its absorption spectrum is used to determine the sample's molecular compositions [6]. Likewise, in our MV-sensitive QPI (MV-QPI), we measure the local changes in the sample-specific optical-path-length delay occurring upon absorption of MIR light in the vicinity of the MV-resonant molecules. This effect, called photothermal effect [7-10], is essentially the change of refractive index of the sample due to increase of temperature as a result of non-radiative decay of the absorbed MIR photon energy. Importantly, with MV-QPI, we can achieve the spatial resolution beyond the MIR diffraction limit due to the visible (VIS) light-based optical-phase-delay detection, maintain high molecular-detection sensitivity based on MIR absorption having ~8 orders of magnitude larger cross-section compared to Raman scattering [8], reduce optical damages to the sample associated with electronic transitions and other nonlinear mechanisms due to wide-field excitation [11], and achieve high-speed imaging ultimately limited by the image sensor's frame rate.

**Principle of MV-QPI**
Figure 1 summarizes the working principle of the constructed QPI system offering sensitivity in the MV fingerprint region. Figure 1(a) sketches the setup designed to operate on semiconductor lasers allowing for easy implementation in various research and application settings. The VIS light source is a nanosecond laser diode (LD) (NPL52C,

Thorlabs) having the optical spectrum with the full-width-at-half-maximum bandwidth of ~2.7 nm peaked at 517.3 nm, the temporal duration of ~130 ns and the pulse energy of 8.25 nJ after beam shaping using a pinhole filter, as shown in Figs. 1(b) and (c). The numerical aperture of the objective lens (LUCPLFLN40X, Olympus) is 0.6, giving the diffraction limited half-pitch resolution of 430 nm. The broadband diffraction phase microscopy technique is used to achieve QPI with high optical-phase-delay sensitivity [12]. The pulsed MIR beam is provided by quantum cascade lasers (QCLs) which is intensity-modulated by a chopper at the rate equivalent to the half harmonic of the image sensor's frame rate. When the chopper is opened, the VIS pulse arrives during or soon after the irradiation of the MIR light such that the optical-path-length delay of the sample under the photothermal excitation is probed, as shown in Fig. 1(c). Generally, the highest photothermal signal is obtained at the falling edge while the highest spatial resolution at the rising edge of the MIR pulse. This pump-probe measurement is repeated several times within one exposure of the image sensor, and the resulting frame is labelled as "MIR ON" image, as shown in Fig. 1(d). When the chopper is closed, the VIS pulse probes the raw optical-path-length delay of the sample without the photothermal excitation, and the resulting frame is labelled as "MIR OFF" frame. Finally, the MV image is obtained by subtracting the MIR ON frame from the MIR OFF frame in order to visualize the change in the optical-path-length delay induced by the photothermal effect, which reflects the MIR absorbance of the sample. We set the pump-probe repetition rate to 1,000 Hz, the image sensor's frame rate 60 Hz with the exposure time of 14 ms and the chopper's rate 30 Hz. In this configuration, 14 pump-probe measurements are integrated in one sensor exposure with the VIS light energy < 115 nJ which is enough to nearly saturate the sensor.

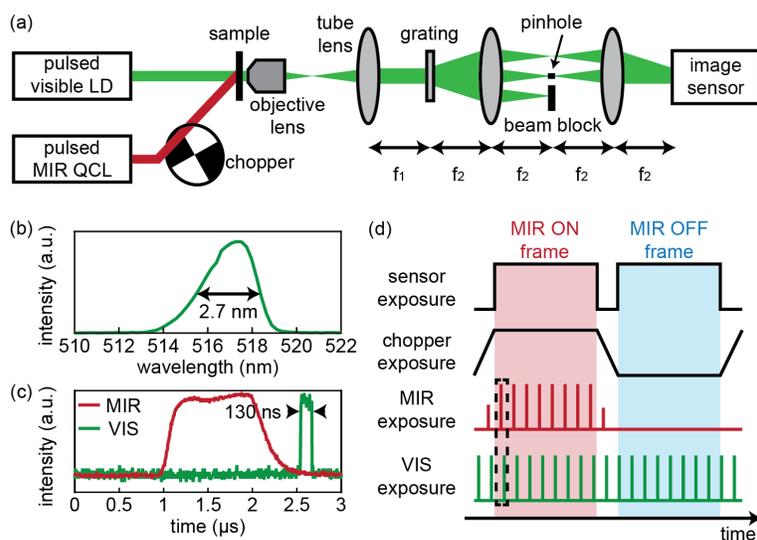

Fig. 1. Principle of MV-QPI. (a) Optical system based on broadband diffraction phase microscopy and semiconductor lasers. The intermediate image of the sample is formed on the grating and in the subsequent 4f relying system, the zeroth order diffraction term is spatially filtered by the pinhole of 25 μm diameter to create a reference wave while only the first order term is allowed to transmit without any alteration, such that the two waves create an interferogram on the image sensor. $f_1$ = 150 mm. $f_2$ = 200 mm. The grating frequency is 145 line-pairs/mm (#66-350, Edmund Optics). The image sensor is acA2440-75um (Basler). (b) Spectral profile of the VIS laser beam. (c) Temporal relation between the MIR and VIS pulses, representing the enlargement of the dashed rectangle shown in (d). (d) Synchronization of the system.

**Characterization of MV-QPI**
We first characterize the basic behaviors of MV-QPI based on an oil sample of a few μm of thickness sandwiching in between $CaF_2$ substrates, which is summarized in Fig. 2. The MIR light source is DO418 (Hedgehog, Daylight Solutions) offering 1 μs MIR pulses ranging between 1,450 – 1,640 $cm^{-1}$. Figure 2(a) shows the characteristic MV image of the sample. The MIR wavenumber is configured to 1,501 $cm^{-1}$ corresponding to one of the absorption peaks of the oil shown in Fig. 2(d) with the pulse energy ~100 nJ. The spatial distribution of the image reflects the local fluence of the focused MIR beam in the sample plane having the peak fluence of ~52 $pJ/\mu m^2$ approximated by the Gaussian beam axes, given by $1/e^2$ contour, of ~47 and ~26 μm. In Fig. 2(b), we observe the temporal evolution of the MV signal at a fixed spatial pixel around the MIR beam center by scanning the time-delay of the VIS probe pulse relative to the MIR excitation pulse. The graph can be fitted by an exponential function with the decay constant of ~200 μs. The thermal decay constant is dependent on the thermal diffusivity and the size of the object with a larger thermal diffusivity and a smaller object size leading to faster thermal decay [13]. The result suggests the pump-probe repetition rate can be set to ~1,000 Hz for observing objects smaller than several tens of micrometers. Since various liquids and polymers have the thermal diffusivity ~$10^{-7}$ m$^2$/s [14,15], similar results could be obtained with other types of material. In Fig. 2(c), we confirm the linear response of the MV signal with respect to the excitation MIR peak fluence at the fixed spatial pixel. Furthermore, in Fig. 2(d), we demonstrate the MIR absorption spectroscopy of the oil sample in the MV fingerprint region with the spectral resolution of 3 cm$^{-1}$. The MV-QPI spectrum is obtained by dividing the MV signal at the fixed spatial pixel by the corresponding MIR peak fluence for each MIR wavenumber. The spectrum shows a good agreement with a reference spectrum obtained by an attenuated-total-reflection Fourier-transform infrared (ATR FTIR) spectrometer (FT/IR-6800, ATR PRO ONE and PKS-D1F, JASCO) with the resolution of 1 cm$^{-1}$. This demonstration validates the quantitative molecular-vibrational spectroscopic performance of our MV-QPI system.

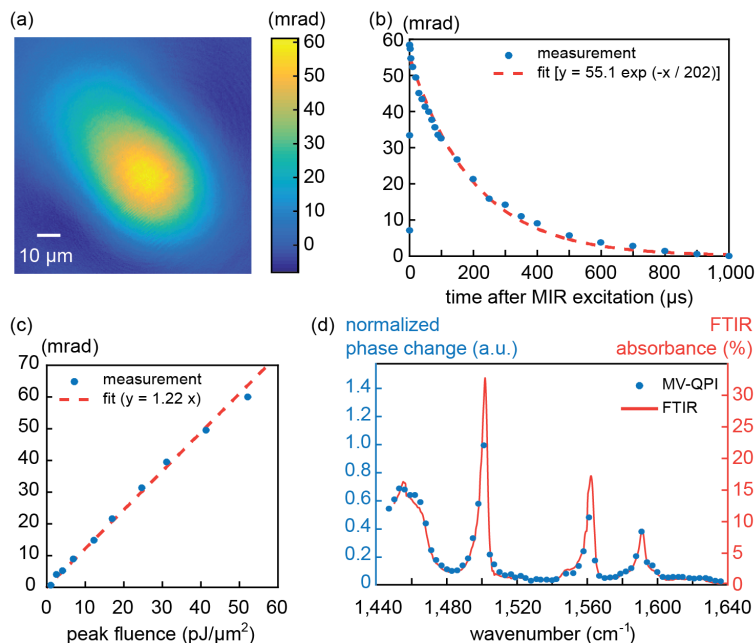

Fig. 2. Characterization of MV-QPI using the oil sample. (a) Typical MV image of the oil reflecting the local fluence of the focused MIR beam in the sample plane. The sample is created by sandwiching a few µL of the liquid oil (index-matching oil with the refractive index of 1.58, Shimadzu) in between two 500 µm-thick CaF$_2$ substrates with 20 mm × 20 mm dimension. (b) MV signal as a function of time delay relative to the 1 µs pulsed MIR excitation. (c) Linear response of the MV signal with excitation MIR peak fluence. (d) MIR absorption spectrum of the oil obtained with our MV-QPI system, showing a good agreement with the reference spectrum obtained by the ATR FTIR spectrometer. The ATR penetration depth is spectrally-normalized and calibrated from ~3.5 µm to ~5 µm by powering the spectrum by 1.5, such that the MV-QPI and the FTIR spectra overlap well with each other. In all the experiments, 1,000 MV images are averaged.

Next we investigate the effect of heat diffusion affecting the spatial resolution of MV-QPI using a 6 µm porous silica bead (43-00-503, Sicastar, micromod Partikeltechnologies GmbH) immersed in index-matching oil (Series A 1.56000, Cargille), as summarized in Fig. 3. We irradiate 10 µs MIR pulses of 1,045 cm$^{-1}$ (QD9500CM1, Thorlabs) resonant to O-Si-O bond of silica and varied the time-delay of the VIS probe pulse with respect to the rising edge of the MIR pulse, such that MV images with effectively different MIR irradiation times are obtained. For the MIR irradiation of 2 – 10 µs the MIR pulse energy is adjusted such that the effective peak MIR fluence is kept to be ~12 pJ/µm$^2$ and 5,000 MV images are averaged, while with the MIR irradiation of 1 µs the peak MIR fluence is ~4 pJ/µm$^2$ hence 20,000 MV images are averaged such that the signal-to-noise ratio (SNR) becomes roughly the same with those in other experimental conditions. Figure 3(a) shows the raw MIR-OFF-state QPI of the bead while Figs. 3(b) – (d) show the MV images obtained with 2, 6, and 10 µs of the MIR irradiation, respectively. We choose to compare the Gaussian radius of each of the obtained bead's images derived using circular Hough transform and indicated by the red dashed circle. In Fig. 3(e), these radii are plotted as a function of the MIR irradiation time. The graph visualizes the degradation of the spatial resolution with a longer MIR irradiation, which can be explained by the heat diffused from the MIR absorber causing temperature-increase and refractive-index-decrease of the peripheral medium [13]. To mitigate this effect, the temporal durations of the MIR and VIS pulses need to be sufficiently shorter than the heat diffusion time of the object, which, in this case, would be several hundreds of nanoseconds.

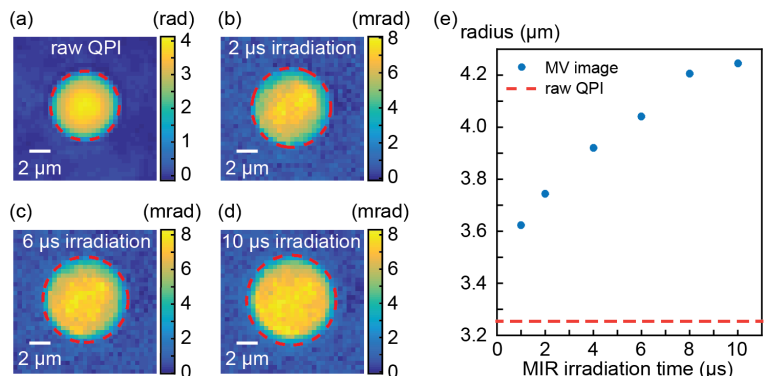

Fig. 3. Degradation of spatial resolution due to heat diffusion effect evaluated using a porous silica bead immersed in index-matching oil. (a) Raw QPI of the bead at MIR OFF state. (b - d) MV image obtained with the MIR irradiation time of (b) 2, (c) 6 and (d) 10 μs. The center coordinate and the Gaussian radius of the bead's contrast are indicated by the dashed red circle in (a - d). (e) Radius of the bead's contrast as a function of the MIR irradiation time. The radius of the raw QPI at MIR OFF state is indicated by the orange dashed line.

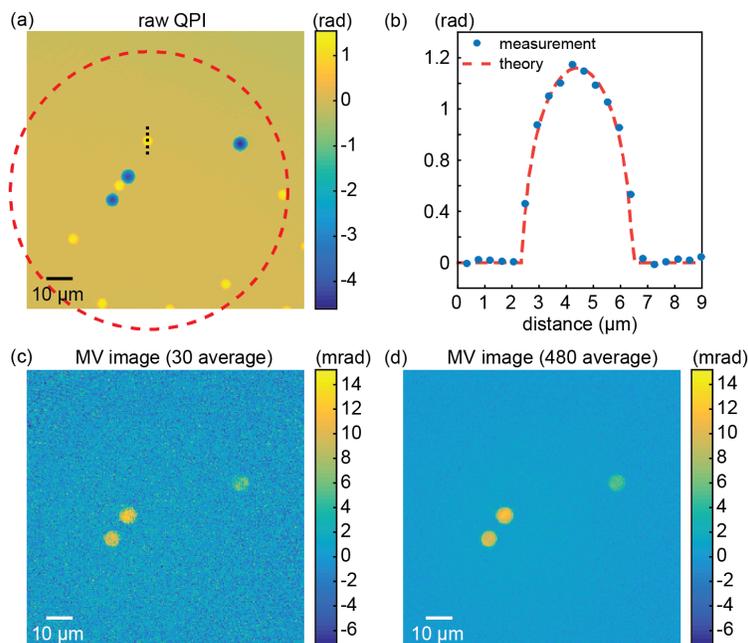

Fig. 4. Wide-field MV-QPI of a mixture of polystyrene and porous silica beads immersed in index-matching oil. (a) Raw QPI of the sample at the MIR OFF state. The wide-field MIR excitation over ~100 μm × 100 μm area (red dashed circle) is realized. (b) Cross-section curve of a polystyrene bead indicated by the black dashed line in (a), showing a good agreement with the theoretical (orange dashed) curve of a 4 μm spherical object. (c - d) MV images of the sample obtained by averaging (c) 30 frames (1 s measurement, SNR 5.0) and (d) 480 frames (16 s measurement, SNR 23.8) in which the silica beads' locations are selectively highlighted. There is a systematic distribution of the MV signal reflecting the local fluence of the MIR beam. The noise floor is given by the standard deviation of 50 × 50 pixels (20 μm × 20 μm) in the top left corner while the signal is given by the mean of 2 × 2 pixels (0.86 μm × 0.86 μm) at the center of the middle silica bead.

**Wide-field MV-QPI of silica-polystyrene beads mixture**
Finally, in Fig. 4, we demonstrate wide-field MV-QPI using a mixture of polystyrene (17135-5, Polybead Microspheres, Polysciences, Inc.) and the porous silica beads immersed in index-matching oil (Series A 1.56000, Cargille). The sample is excited by 5 μs MIR pulses of 1,045 cm$^{-1}$ with the pulse energy ~100 nJ (QD9500CM1, Thorlabs). Figure 4(a) shows the raw QPI of the mixture sample at the MIR OFF state. The polystyrene and the silica beads, respectively, show positive and negative contrasts in the QPI as their refractive indices are higher (1.600) and lower (1.461) than that of the oil (1.569). We realize the wide-field excitation over ~100 μm × 100 μm area of the Gaussian beam diameter, resulting in the peak fluence of ~25 pJ/μm$^2$. The cross-section curve of one of the polystyrene beads indicated by the black dashed line in the image is plotted in Fig. 4(b), which shows a good agreement with the theoretical curve of a spherical object with a diameter of 4 μm, validating the QPI capability of our system. Figure 4(c) shows the MV image of the sample with the averaging number of 30, corresponding to 1 s measurement and offering the SNR of 5.0, in which the locations of the silica beads are selectively highlighted. The higher SNR can be achieved with a longer measurement time as shown in Fig. 4(d), where the MV image is obtained with the averaging number of 480, corresponding to 16 s measurement and offering the SNR of 23.8. Overall, this demonstration verifies the wide-field MV-QPI capability of our technique.

**Discussions and conclusions**
Compared to other high-speed label-free chemical imaging methods, our constructed MV-QPI system excels in its reduced optical damages to the sample and simple optical implementation. In coherent Raman imaging [16,17], for example, intense optical pulses of ps – fs duration and ~1 – 10 nJ pulse energy are tightly focused to a nearly-diffraction-limited spatial point of ~1 μm × 1 μm in the VIS to the near-infrared regime, resulting in the fluence of ~1 – 10 nJ/μm$^2$ which could induce sample damages via electronic and other nonlinear transitions [11]. In our wide-field MV-QPI, on the other hand, the VIS and MIR fluence is ~10 pJ/μm$^2$ which is 2 – 3 orders of magnitude lower, yet it achieves the moderate imaging speed of 1 frame per second. The possible sample damage is associated with local temperature change in the vicinity of the MIR absorber, which, in our demonstration with the silica beads, is estimated to be ~0.4 K, as the temperature dependence of the oil's refractive index is -4.26 × 10$^{-4}$ /K. Furthermore, our MV-QPI operates on semiconductor lasers and does not require any sophisticated devices such as solid-state oscillators, galvanometer scanners, acousto-optic modulators, etc., ensuring easy implementation in other research or application settings.

The current sensitivity and spatial resolution of our MV-QPI are, respectively, limited by the energy and duration of the MIR pulse source which are in the order of 100 nJ/pulse and 1 μs. A higher pulse energy can produce larger change of the optical-path-length delay resulting in a higher SNR, while a shorter pulse width can reduce the heat diffusion effect that occurs within the pulse duration. Therefore, the ideal MIR light source would be a kilohertz laser having 10 – 100 μJ pulse energy and 10 - 100 ns pulse duration which could boost the image acquisition rate by 2 - 3 orders of magnitude. Such a high frame rate is, in principle, realizable with MV-QPI due to the wide-field parallelized detection scheme of MIR photothermal signal with a two-dimensional image sensor, and beyond the reach of other state-of-the-art label-free molecular-sensitive microscopes, such as coherent Raman [16,17] and other MIR photothermal [7-10] microscopes, which employ a point-scanning mechanism for image acquisition.
We also note that the label-free, simultaneous and in-situ acquisition of the quantitative morphology and the MV absorption contrast is a unique capability of our MV-QPI which could promote new findings in studies on complex dynamics of an optically-transparent system. This is not readily achieved if an existing wide-field QPI technique is used in combination with other existing chemical imaging technique, as the use of two different detectors and/or switching between the two imaging modes would be required. Our MV-QPI could enable the cross-correlative analysis without sacrificing spatial or temporal consistency between the two complementary contrasts of the microscopic specimen.

To conclude, we have proposed and demonstrated the wide-field label-free MV-fingerprinting microscopy operating on semiconductor lasers in the framework of QPI utilizing the MIR photothermal effect. We have shown that our MV-QPI system offers the MIR absorption spectroscopic performance comparable to that of the conventional FTIR spectrometer and demonstrated wide-field label-free MV-QPI over ~100 μm × 100 μm area based on the silica-polystyrene beads mixture at the rate of 1 frame per second with the visible-light-based spatial resolution of 430 nm. Our MV-QPI allows for label-free chemical imaging with an exceptionally reduced optical damages to the sample, as the fluence is in the order of 10 pJ/μm$^2$ which is 2 – 3 orders of magnitude lower than that of other high-speed label-free chemical imaging methods such as coherent Raman imaging. The imaging speed of MV-QPI can be increased by using a high-energy nanosecond kilohertz MIR pulse source up to the rate ultimately limited by the frame rate of the image sensor, which can be beyond the reach of other state-of-the-art label-free molecular-sensitive imaging techniques which are based on a point-scanning mechanism for image synthesis. With such an improved system, we expect our MV-QPI could offer new insights for optically-transparent complex dynamics by enabling high-speed and label-free cross-correlative analysis based on the quantitative morphology and the MV absorption contrast of microscopic targets with accurate spatial and temporal consistencies.

**Funding**
JST PRESTO (JPMJPR17G2); JSPS KAKENHI (17H04852, 17K19071); Research Foundation for Opto-Science and Technology; The Murata Science Foundation.

**Acknowledgment**
We thank Makoto Kuwata-Gonokami and Junji Yumoto for letting us use their equipment.